\documentclass[prb,superscriptaddress,twocolumn,showkeys]{revtex4-2}

\usepackage{graphicx}
\usepackage{hyperref}
\usepackage{amsmath}
\usepackage{amssymb}
\usepackage{bm}
\usepackage{units}
\usepackage{xcolor}
\usepackage{ulem}

\usepackage{amsfonts}
\usepackage{tabularx, ragged2e}
\usepackage{tcolorbox}
\usepackage{mathptmx}
\usepackage{subfiles}
\usepackage{multirow}
\usepackage{mathtools}
\usepackage{float}

\newcommand{\ii}{\iota}
\newcommand{\bra}[1]{\left< #1 \right|}
\newcommand{\ket}[1]{\left| #1 \right>}

\newcommand{\TO}{\begin{tcolorbox}[width=\textwidth,colback={green}]}
	\newcommand{\DO}{\end{tcolorbox}}
\definecolor{muschi}{cmyk}{0,.2,0,0} 
\newcommand{\AT}{\begin{tcolorbox}[width=\textwidth,colback={muschi}]}
	\newcommand{\TA}{\end{tcolorbox}}
\definecolor{blut}{cmyk}{0,.4,0,0} 
\newcommand{\PRO}{\begin{tcolorbox}[width=\textwidth,colback={blut}]}
	\newcommand{\BLEM}{\end{tcolorbox}}
\definecolor{todo}{cmyk}{0,0,0,.45}

\definecolor{boldi}{cmyk}{.9,.1,0,.55} 

\definecolor{boldy}{cmyk}{0,.9,.5,.25}

\newcommand{\scale}{0.7}

\newcommand{\lcpq}{Univ. Toulouse, CNRS, LCPQ, and European Theoretical Spectroscopy Facility (ETSF), Toulouse, France}
\newcommand{\lpt}{Univ. Toulouse, CNRS, LPT, and European Theoretical Spectroscopy Facility (ETSF), Toulouse, France}

\begin{document}
	
 \title{Ground and excited-state properties of the extended Hubbard dimer from the multichannel Dyson equation}
	
	\author{Stefano Paggi}
 \email{spaggi@irsamc.ups-tlse.fr}
	\affiliation{\lpt}
		\author{J. Arjan Berger}
	\email{arjan.berger@irsamc.ups-tlse.fr}
	\affiliation{\lcpq}
 \author{Pina Romaniello}
 \email{pina.romaniello@irsamc.ups-tlse.fr}
	\affiliation{\lpt}

	\begin{abstract}
We have recently presented the multichannel Dyson equation as an alternative to the standard single-channel Dyson equation. 
While the latter involves a single many-body Green's function, the former uses a multichannel Green's function in which two or more many-body Green's functions are coupled.
Quasiparticles and satellites are thus naturally treated on equal footing in the multichannel Dyson equation.
To assess the accuracy of our approach we apply it here to the ground- and excited-state properties of the extended Hubbard dimer, an exactly solvable model for H$_2$.
In particular, we focus on the potential energy surface as well as the corresponding spectral functions and HOMO-LUMO gaps, which are well-known challenges for many-body approximations such as second Born and $GW$. 
We show that the multichannel Dyson equation gives overall very good results for all properties considered and outperforms both $GW$ and second Born.
In particular, the multichannel Dyson equation yields the correct ground-state energy and HOMO-LUMO gap in the dissociation limit contrary to $GW$.
 \end{abstract}
	
	\maketitle

\section{Introduction}
The single-channel Dyson equation for the 1-body Green's function (1-GF) has been successfully used for decades to simulate photoemission spectra.
It relates an independent-particle 1-GF, e.g., obtained from Hartree-Fock (HF) or Kohn-Sham DFT, to the 1-GF of the interacting system through an effective potential, the 1-body self-energy (1-SE). \cite{luciaReining}
However, with currently available approximations to the 1-SE, e.g., $GW$,\cite{Hed65,Hybertsen_1986,Hed99,Ary98} the success of the single-channel Dyson equation is generally limited to the coherent part of photoemission spectra, i.e., quasiparticles (QPs), for which typically a good agreement with experiment is obtained for weakly correlated systems.
Instead, the incoherent part of the spectra, i.e., the satellites, does generally not compare well to that observed in experiment (see e.g. ~\onlinecite{Ary92,Ary96,Vos_1999,Kheifets_PRB03,Guz11,Dis21-2,Dis23-1}). 
The main reason for the different precision with which quasiparticles and satellites are calculated within the single-channel Dyson equation is that they are not treated on equal footing.~\cite{Riv22,riva_prl,riva_prb}
While approximate quasi-particle energies are already present in the independent-particle 1-GF, the corresponding satellite energies are absent.
As a consequence, the exact 1-SE is a complicated function of the energy which current approximations fail to capture.

For this reason, we recently introduced the multi-channel Dyson equation (MCDE) which couples various $n$-body Hartree-Fock Green's function through a multichannel self-energy.~\cite{riva_prl,riva_prb}
For example, we showed that coupling the $1$-body HF Green's function to the hole-hole-electron ($2h1e$) and electron-electron-hole ($2e1h$) channels of the $3$-body HF Green's function, quasiparticles and 3-particle satellites are put on equal footing.
Multichannel Dyson equations naturally split into two classes, those involving only $n$-GFs with $n$ even and those involving $n$-GFs with $n$ odd.\cite{riva_prb_25}
The various MCDE can be further distinguished by the difference in the number of electrons in the initial and final states, denoted $s$, and the highest-order Green's function that is involved in the MCDE, denoted $n$. The corresponding MCDE is then denoted $(n,s)$-MCDE. For example, the MCDE mentioned above that couples
the $1$-body HF Green's function to the $2h1e$ and $2e1h$ channels of the $3$-body HF Green's function is the $(3,1)$-MCDE.

Moreover, we proposed a systematic approach to construct approximations for multi-channel self-energies.
These approximations yield static multi-channel self-energies.
Therefore, each MCDE can be rewritten as an eigenvalue problem with an effective Hamiltonian, which can be solved using standard numerical techniques.~\cite{Hay72,Schm03,Her05}
Moreover, a static self-energy guarantees the absence of unphysical solutions.~\cite{Lan12,Ber14,Sta15,Tar17}
We showed that, within the $(3,1)$-MCDE, this approach yields exact results for the spectral function of the Hubbard dimer at 1/4 and 1/2 filling and at any interaction strength \cite{riva_prl}, 
demonstrating the potential of the $(3,1)$-MCDE to correctly describe satellites in the spectral functions of real molecules and materials even at strong correlation.
In order to conclude that our results are exact, it was crucial to use a model that can be solved analytically,  such as the Hubbard dimer. 

Unfortunately, the Hubbard dimer does not resemble a real homo-nuclear dimer such as H$_2$ because its corresponding Hamiltonian only contains interactions between electrons occupying the same atomic orbital.
However, in order to correctly simulate the potential energy curve of a dimer the electrons should also interact when they occupy atomic orbitals centered on different atoms.
Therefore, in this work we will apply the MCDE to the extended Hubbard dimer whose Hamiltonian also contains an interaction between electrons when the two electrons occupy orbitals on different atoms and, crucially, can still be solved analytically.~\cite{Giesbertz_2018,Lani_2024}
This will allow us to investigate potential improvements of our approximation to the 3-SE.

The paper is organized as follows.
In section \ref{sec:theory}, we provide the theoretical details of the $(3,1)$-MCDE and the corresponding approximation to the 3-body self-energy.
We also report the standard second Born (2B) and $GW$ approximations which will be used to assess the accuracy of the $(3,1)$-MCDE.
In Sec.~\ref{sec:illustration}, we present and discuss the results obtained with the $(3,1)$-MCDE for the potential energy surface, spectral functions and HOMO-LUMO gaps of the extended Hubbard dimer. 
The MCDE results are compared to the exact results, and those obtained with the 2B and $GW$ approximations. 
Finally, we draw conclusions and provide an outlook in Sec.~\ref{sec:conclusions}
\section{Theoretical framework\label{sec:theory}}
In this section we will first discuss the standard single-channel Dyson equation followed by a discussion of our multichannel Dyson equation.
When possible, we suppress the space-spin variables in the following for notational convenience. Moreover, we will suppress the subscript of an $n$-body Green’s function $G_n$ whenever the number of indices of the Green’s function are explicitly given, e.g., $G_{3 ijl;mok}\rightarrow G_{ijl;mok}$.
We use atomic units $\hbar=m=e=1$ and work at zero temperature throughout the paper.
\subsection{Single-channel Dyson equation}
The 1-GF is commonly obtained by solving the following single-channel Dyson equation
\begin{equation}
\label{Eq:Dysonsingle}
G_{1}(\omega)=G^0_1(\omega)+G^0_1(\omega)\Sigma_1[G_1](\omega)G_1(\omega),
\end{equation}
where $G^0_1(\omega)$ is the noninteracting 1-GF and $\Sigma_1(\omega)$ is the one-body self-energy which is a functional of the 1-GF.
It can be decomposed as
\begin{equation}
\label{Eq:Sigma1decomp}
\Sigma_1(\omega)=  \Sigma^{\text{HF}}_1[G_1]+  \Sigma^{\text{c}}_1[G_1](\omega)
\end{equation}
with $\Sigma^{\text{HF}}_1$ is the static Hartree-Fock contribution to the self-energy and $\Sigma^{\text{c}}_1(\omega)$ the dynamical correlation self-energy.\cite{Sch78,Sch83,Nie84,vonBarth_PRB96}
The single-channel Dyson equation can thus be rewritten as 
\begin{equation}
G_{1}(\omega)=G^{\text{HF}}_1(\omega)+G^{\text{HF}}_1(\omega)\Sigma^{\text{c}}_1[G_1](\omega)G_1(\omega).
\label{Eq:Dysoncorr}
\end{equation}
%
In practice, $\Sigma^{\text{c}}_1(\omega)$ has to be approximated to allow for calculations on many-body systems.
Two well-established approximations to $\Sigma^{\text{c}}_1(\omega)$ are the second Born and $GW$ approximations.
In orbital space they are given by, respectively,
\begin{align}
\Sigma^{\text{c},2B}_{i;m}(\omega)&=\sum_{jklnop}v_{ipko}\bar{v}^*_{njlm} \iint\frac{d\Omega d\omega'}{(2\pi)^2} \nonumber\\
&\times  G_{o;n}(\omega-\Omega)G_{k;j}(\omega'+\Omega)G_{l;p}(\omega')\\
  \Sigma^{\text{c},GW}_{i;m}(\omega)&=\ii\sum_{jl}\int\frac{d\omega'}{2\pi}G_{j;l}(\omega+\omega')W^c_{ilmj}(\omega')  
\end{align}
where $\bar{v}_{ijkl}={v}_{ijkl}-{v}_{ijlk}$.
The matrix elements of the Coulomb potential $v_c$ and the correlation part $W^c=W-v_c$ of the screened Coulomb potential $W$ are given by
\begin{align}
v_{ijkl} & =\iint \mbox{d}x_1 \mbox{d}x_2 \phi^\ast_i (x_1)\phi_j^\ast (x_2)v_c(\textbf{r}_1,\textbf{r}_2)\phi_k(x_2)\phi_l(x_1),
\\
W^c_{ilmj} & =\int \mbox{d}x_1 \mbox{d}x_2 \phi^\ast_i (x_1)\phi_l^\ast (x_2)W^c(\textbf{r}_1,\textbf{r}_2)\phi_m(x_2)\phi_j(x_1),
\end{align}
in which $x_1$ and $x_2$ are combined space-spin variables.
These approximations depend themselves on the 1-GF which is the quantity we want to obtain by solving Eq.~\eqref{Eq:Dysonsingle}.
Therefore, in principle, the single-channel Dyson equation in Eq.~\eqref{Eq:Dysonsingle} and $\Sigma^{\text{c,2B}}_1$ (or $\Sigma^{\text{c},GW}_1$) should be solved self-consistently.
However, in many calculations, only a single iteration of these equations is performed, e.g., for the 2B approximation one typically solves
\begin{equation}
\label{Eq:oneshot2B}
G_{1}(\omega)=G^{\text{HF}}_1(\omega)+G^{\text{HF}}_1(\omega)\Sigma^{\text{c,2B}}_1[G_1^{\text{HF}}](\omega)G_1(\omega).
\end{equation}
We have a similar expression for the single-particle Dyson equation involving the $GW$ self-energy.
This is also the approach we will adopt in this work.
\subsection{Multichannel Dyson equation}

In a similar way as for the single-particle Dyson equation in Eq.~\eqref{Eq:Dysoncorr} 
we can set up a Dyson equation for $G_3(\omega)$ which contains both the $2h1e$ and $2e1h$ channels of the 3-GF according to
\begin{equation}
	\label{eq:gDyson}
	G_3(\omega)=G^{\text{HF}}_{3}(\omega)+G^{\text{HF}}_{3}(\omega)\Sigma^c_3(\omega)G_3(\omega),
\end{equation}
where $G^{\text{HF}}_{3}(\omega)$ is the corresponding 3-GF in the HF approximation
and $\Sigma^c_3$ denotes the corresponding three-body correlation self-energy. 

It is convenient to reformulate Eq.~\eqref{eq:gDyson} in the basis of the one-particle HF spin-orbitals ${\phi^{\text{HF}}_i}$ that diagonalize $G_{3}^{\text{HF}}$. 
Within this basis, $G_{3}^{\text{HF}}$ naturally decomposes into a one-particle channel $G^{\text{HF}}$ and a three-particle channel $G^{\text{HF},3\text p}$, leading to the following block-diagonal structure,
\begin{equation}
	\label{eq:gpartition}
	G_{3}^{\text{HF}}(\omega)=\left(
	\begin{array}{cc}
		G^{\text{HF}}(\omega)&0\\
		0&G^{\text{HF},3\text{p}}(\omega)\\
	\end{array}
	\right),
\end{equation}
where
\begin{align}
	\label{eq:g01im}
	G^{\text{HF}}_{i;m}(\omega)&=\frac{\delta_{im}}{\omega-\epsilon^{\text{HF}}_i+\ii \eta \text{sgn}(\epsilon_i^{\text{HF}}-\mu)},\\
	\label{eq:g03im}
	G^{\text{HF},3\text{p}}_{i>jl;m>ok}(\omega)&=\frac{\delta_{im}\delta_{jo}\delta_{lk}(f_i-f_l)(f_j-f_l)}{\omega-\epsilon^{\text{HF}}_i-(\epsilon^{\text{HF}}_j-\epsilon^{\text{HF}}_l)+\ii \eta \text{sgn}(\epsilon_i^{\text{HF}}-\mu)}.
\end{align}
In these expressions, $\epsilon^{\text{HF}}_i$ and $f_i$ are the HF orbital energies and occupation numbers, respectively, where $f_i=1$ for $\epsilon_i \leq \mu$ and 0 otherwise with $\mu$ being the chemical potential. 
We note that the conditions $i>j$ and $m>o$ in Eq.~ (\ref{eq:g03im}) prevent double counting, while $(f_i-f_l)(f_j-f_l)$ ensures that $G_{3}^{\text{HF}}$ is restricted to its $2h1e$ and $2e1h$ channels.
We note that the poles of $G^{\text{HF}}$ correspond to approximate QP energies, while the poles of $G^{\text{HF},3\text p}$ correspond to approximate satellite energies.

Expressing Eq.~\eqref{eq:gDyson} in the basis that diagonalizes $G^{\text{HF}}(\omega)$ we obtain a multichannel Dyson equation in which the self-energy can be expressed as follows

\begin{equation}
	\Sigma^c_3=
	\left(
	\begin{array}{cc}
		\Sigma^{1\text{p}} & \Sigma^{1\text{p}/3\text{p}}\\
		{\Sigma}^{3\text{p}/1\text{p}} & \Sigma^{3\text{p}}
	\end{array}
	\right).
 \label{Eqn:S3}
\end{equation}
where $\Sigma^{1\text{p}}$ and $\Sigma^{3\text{p}}$ represent the self-energies of the 1-particle and 3-particle channels, respectively, 
while $\Sigma^{1\text{p}/3\text{p}}$ and $\Sigma^{3\text{p}/1\text{p}}$ couple the two channels.
To make practical calculations feasible, an approximation for the $(3,1)$-MCDE self-energy $\Sigma_3$ is required.
We approximate $\Sigma_3$ by including all interactions that are of first order in the Coulomb interaction.
However, all contributions to $G_1(\omega)$ that are of first-order in the interaction are already contained in $G^{\text{HF}}(\omega)$.
Therefore, $\Sigma^{1\text{p}}=0$ to avoid the double counting of correlation. 
We note that this implies that $G^{\text{HF}}_1$ should be calculated from $\Sigma^{HF}[G_1]$ which would require a self-consistent calculation. In practice we avoid this and use $\Sigma^{HF}[G^{HF}_1]$ to obtain $G^{\text{HF}}_1$.
These considerations lead to the following static expression for the self-energy,
\begin{align}
	\Sigma^{3 \text p}_{ijl;mok}&=\!\! [(1\!-\!f_i)\!(1\!-\!f_j)f_l\!-\!f_if_j(1\!-\!f_l)][\delta_{lk} \bar v_{ijom} \nonumber \\ &+\!\delta_{mj}\bar v_{iklo} \!+\! \delta_{io} \bar v_{jklm} \!-\! \delta_{oj}  \bar v_{iklm} \!-\! \delta_{im}  \bar v_{jklo}] ,\label{selfthird:eq}\\
	\Sigma^{1\text{p}/3\text{p}}_{i;mok}&=\bar v_{ikom}, \label{selfcoupling:eq} \\
	 \Sigma^{3\text{p}/1\text{p}}_{ijl;m}&=\bar v_{ijlm}, \label{selfcouplingtilde:eq}\\
	\Sigma^{1 \text p}_{i;m}&=0. \label{selfone:eq}
\end{align}
%

To obtain an equation that can be solved in practice using standard numerical methods, we map Eq.~\eqref{eq:gDyson} onto an effective three-particle equation.~\cite{riva_prb}
We hence obtain
\begin{equation}
	\label{eq:g3effHammy}
	G_{ijl;mok}=\left[\omega I - H_3^{\text{eff}}\right]^{-1}_{ijl;mok},
\end{equation}
in which the effective Hamiltonian $H_3^{\text{eff}}$ is given by
\begin{equation}
	H_3^{\text{eff}}=\left(
	\begin{array}{cc}
		H^{1\text p} & {H}^{\text{1p/3p}} \\
	H^{\text{3p/1p}} & H^{3\text p}
	\end{array}
	\right),
\end{equation}
where 
\begin{align}
    H^{1\text p}_{i;m}&=  \epsilon^{\text{HF}}_i \delta_{im}
    \\
    H^{\text{1p/3p}}_{i;m>ok}&= (f_{m}-f_{k}) (f_{o}-f_{k}) \Sigma^{\text{1p/3p}}_{i;mok},
    \\
   H^{\text{3p/1p}}_{i>jl;m}& = (f_{i}-f_{l})(f_{j}-f_{l})\Sigma^{\text{3p/1p}}_{ijl;m},
    \\
    H^{3\text p}_{i>jl;m>ok}&= (\epsilon^{\text{HF}}_{i}-(\epsilon^{\text{HF}}_{l}-\epsilon^{\text{HF}}_{j})) \delta_{im}\delta_{jo}\delta_{lk}\nonumber\\
    &+ (f_{i}-f_{l}) (f_{j}-f_{l}) \Sigma_{ijl;mok}^{3p}.
\end{align}
Since $\Sigma_3^c$ is static in our approximation the effective Hamiltonian in Eq.~\eqref{eq:g3effHammy} is static as well.
Therefore, the 3-GF can be reconstructed from the eigenvectors and eigenvalues of $H^{\text{eff}}$.
In particular, the 1-GF, which corresponds to the head of the 3-GF, can be obtained as
\begin{align}
\label{Eqn:G1_MCDE}
    G_{i;m}(\omega)& =  \sum_{\lambda}\frac{A^i_{\lambda}A^{*m}_{\lambda}}{\omega-E_{\lambda}},
    \end{align}
where $E_{\lambda}$ and $A_{\lambda}$ are the eigenvectors and eigenvalues of $H_3^{\text{eff}}$, respectively.

Finally, we point out that the 1-GF within the 2B approximation obtained from Eq.~\eqref{Eq:oneshot2B} can also be obtained from the ($3,1$)-MCDE.
The 2B 1-GF is equal to the 1-particle channel of the the 3-GF when $\Sigma^{3 \text p}_{ijl;mok}=0$ ~\cite{riva_prb}.
In our formulation this corresponds to the head of the 3-GF.
The results obtained within the 2B approximation presented in the next section were obtained within our multichannel formulation.
%
%
\subsection{Properties}
As mentioned before, in this work we will focus on the potential energy surface and the corresponding spectral functions and HOMO-LUMO gaps of an H$_2$ model.
The ground-state energy $E_0$ can be calculated from the 1-GF using the Galitskii-Migdal (GM) energy functional which is given by
\begin{equation}
\label{Eq:GM}
 E_0 = \frac{1}{2} \sum_{ij} \int\frac{\text{d}\omega }{2\pi\ii}e^{i\omega0^+}\left[\omega \delta_{ij}+h_{i;j}\right]G_{j;i}(\omega),
\end{equation}
%
where $h_{i;j}$ are the one-particle matrix elements.
The spectral function can be obtained from the 1-GF according to
\begin{equation}
 A(\omega)=\frac{1}{\pi}\sum_i\left|\mbox{Im}G_{(i;i)}(\omega)\right|
\end{equation}
where $G_{(i;i)}$ are the diagonal components of the 1-GF.
Finally, the HOMO-LUMO gap is simply the difference of the two QP energies.
%
%
\section{Application to the extended Hubbard dimer\label{sec:illustration}}
The Hamiltonian of the extended Hubbard dimer (EHD) is given by
%
%
\begin{align}
	\label{eq:marzipan_hubbard}
	\hat{H}^{\text{EHD}}&=\epsilon_0 \sum_{\sigma;i=1,2}\hat{c}^\dagger_{i\sigma}\hat{c}_{i\sigma}-t\sum_{\sigma}\sum_{\substack{i,j=1,2 \\ i\neq j}}\hat{c}^\dagger_{i\sigma}\hat{c}_{j\sigma}
	\nonumber\\
	&+\frac{U}{2}\sum_{\substack{\sigma\neq\sigma^\prime \\ i=1,2}}\hat{c}^\dagger_{i\sigma}\hat{c}^\dagger_{i\sigma^\prime}
	\hat{c}_{i\sigma^\prime}\hat{c}_{i\sigma}
	+ \frac{w}{2}\sum_{\substack{\sigma, \sigma^\prime \\ i,j=1,2\\i\neq j}}\hat{c}^\dagger_{i\sigma}\hat{c}^\dagger_{j\sigma^\prime}
	\hat{c}_{j\sigma^\prime}\hat{c}_{i\sigma},
\end{align}
where $\hat{c}^\dagger_{i\sigma}$ ($\hat{c}_{i\sigma}$) is the creation (annihilation) operator corresponding to the orbital $i=1,2$ and spin $\sigma=\uparrow,\downarrow$,
$\epsilon_0$ is the onsite energy, $t$ the hopping parameter, $U$ the onsite Coulomb interaction, and
$w$ the intersite Coulomb interaction. 
The dimer has two bonding spin-orbitals, and two antibonding spin-orbitals.
In this paper we focus on the half-filling case, e.g., the two-electron case, as it resembles the H$_2$ molecule.  

\begin{figure}[t]
\includegraphics[scale=\scale]{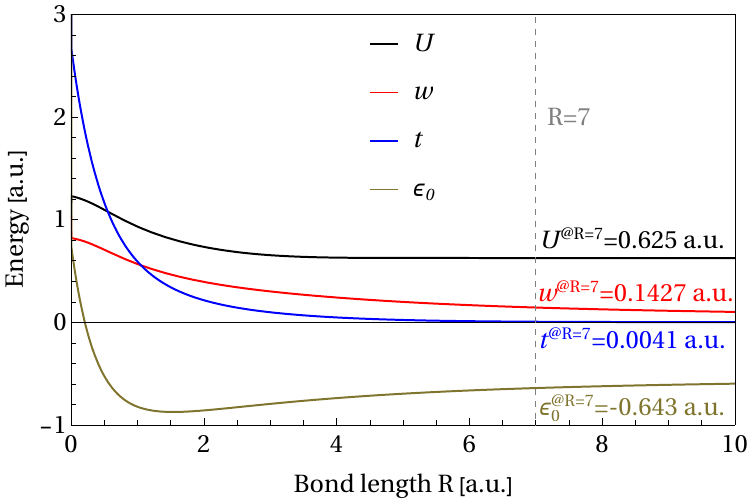}
\caption{The evolution of the optimised parameters of the extended Hubbard dimer as a function of the bond length $R$. The values of the parameters corresponding to $R=7$ a.u. are given explicitly.}
\label{fig:bondlengthRparaDep}
\end{figure}

Giesbertz \textit{et al.} showed that by appropriately choosing the values of the four parameters in the extended Hubbard dimer the potential energy surface of the H$_2$ molecule calculated within a minimal basis set could be accurately reproduced \cite{Giesbertz_2018}.
It is instructive to investigate the evolution of the optimised parameters $\epsilon_0$, $t$, $U$, and $w$ as a function of the bond distance $R$.
This was done by Lani \textit{et al} in Ref.~\onlinecite{Lani_2024}.
To keep this paper self-contained we report a similar analysis in Fig. \ref{fig:bondlengthRparaDep}.
As expected both $t$ and $w$ vanish in the dissociation limit while $U$ remains finite in this limit.
\subsection{Exact results}
We obtained analytical expressions of the exact $2h1e$ and $1e2h$ channels of the 3-GF corresponding to the extended Hubbard dimer from its Lehmann representation.~\cite{Riv22}
Its calculation at half-filling thus requires the eigenvalues and eigenvectors of the EHD Hamiltonian given in Eq.~\eqref{eq:marzipan_hubbard} for $N=1,2,3$ with $N$ the number of electrons.
All details are given in Appendix \ref{App:exact}.
The exact electronic energy of the ground-state of the EHD is equal to the lowest eigenvalue of the EHD Hamiltonian for $N=2$.
It is given by
\begin{align}
E_0^{\text{EHD}}&=2\epsilon_0 + \frac{U+w-\sqrt{16t^2+(U-w)^2}}{2}.
\end{align}
We note that for $w=0$ one retrieves the total energy of the standard Hubbard dimer (HD) as one should.
It is evident that
\begin{equation}
\label{Eq:linkEHD-HD}
E_0^{\text{EHD}} = w+\lim_{ U\to U-w} E_0^{\text{HD}},
\end{equation}
and similar expressions hold for all eigenvalues of $\hat{H}^{EHD}$ as has already been pointed out in Ref.~\onlinecite{Lani_2024}. 
We will come back to this link between the standard and extended Hubbard dimers in the next subsection.

\begin{figure}[t]
\centering
\includegraphics[scale=\scale]{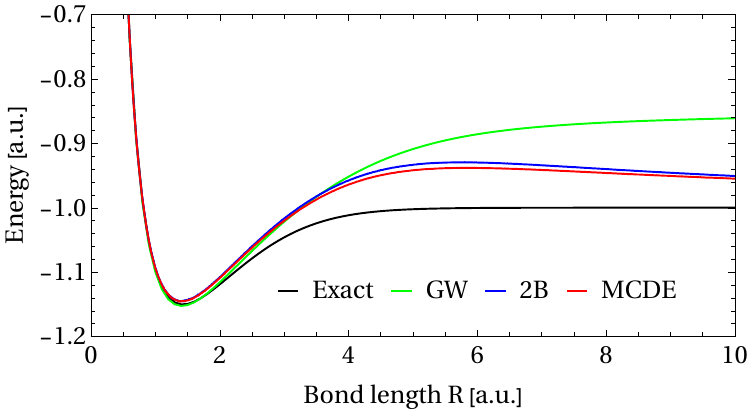}
\caption{The ground-state energy as a function of the bond length $R$ of the extended Hubbard dimer obtained within various approximations (MCDE, 2B, and $GW$) compared to the exact potential energy surface.}
\label{fig:extHubbard_gal_migdal_multiplot_3}
\end{figure}

\begin{figure*}[t]
\centering
\includegraphics[scale=\scale]{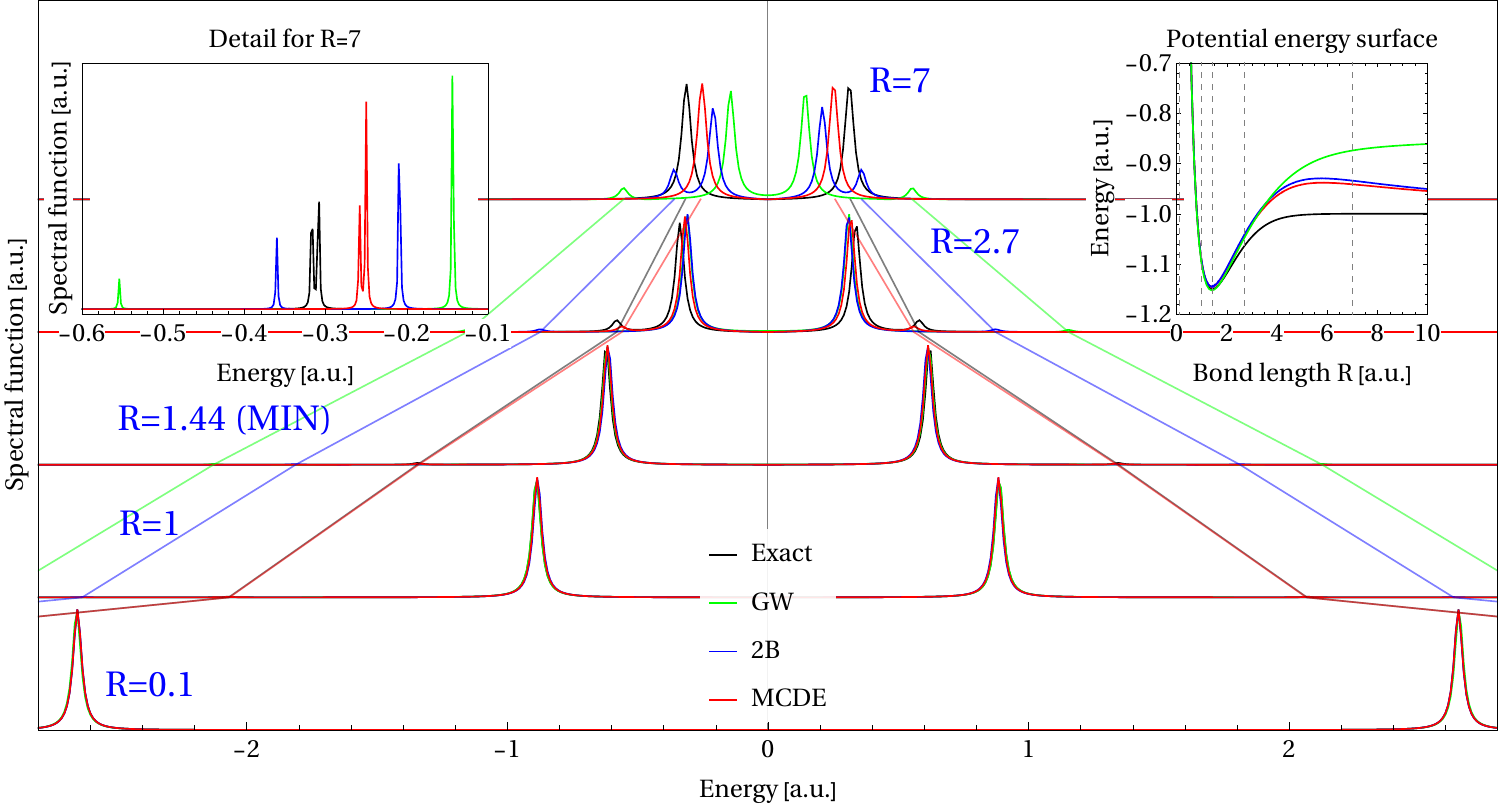}
\caption{Photoemission spectra $A(\omega)$ for various values of the bond length $R$ of the extended Hubbard dimer obtained within various approximations (MCDE, 2B, and $GW$) compared to the exact spectra. 
The broadening $\eta=0.02$ a.u.. The two lines on the left and right of the QP peaks showcase the positions of the satellites. The right inset displays the corresponding potential energy surfaces, with vertical gray lines marking the $R$-values for which the spectral functions are reported. The left inset displays a detail of the spectral function at $R=7$ a.u. in which $\eta=0.001$ a.u..}
\label{fig:extHubbard_G1G3_2B_Exqp_Exact_H2}
\end{figure*}

\begin{figure}[t]
		\centering
		\includegraphics[scale=\scale]{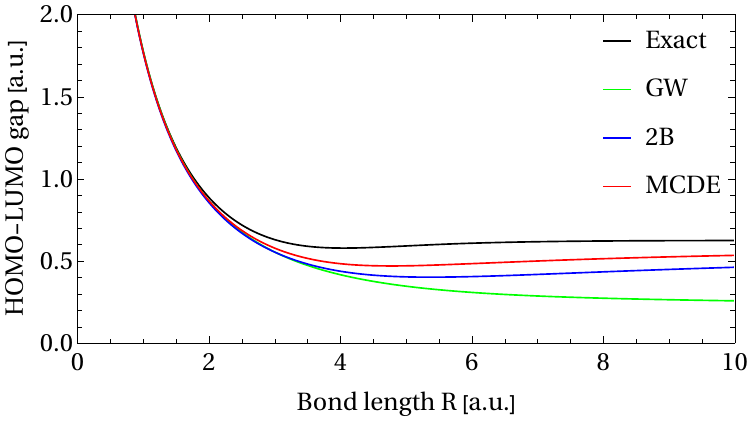}
		\caption{The HOMO-LUMO gap $\Delta_g$ as a function of the bond length $R$ of the extended Hubbard dimer obtained within various approximations (MCDE, 2B, and $GW$) compared to the exact gap.}
		\label{Fig:bandgaps}
	\end{figure}

Adding the $1/R$ nuclear repulsion energy to the electronic energy, where $R$ is the intersite distance, results in the potential energy surface (PES), which we report in Fig. \ref{fig:extHubbard_gal_migdal_multiplot_3}. 
One can observe its resemblance to the PES of the H$_2$ molecule.
Of particular interest is the dissociation limit, where strong electron correlation comes into play, which is a regime often difficult to describe with approximate self-energies.
In this limit $\epsilon_0=-1/2$ a.u., $t=w=0$ a.u., and $U=5/8$ a.u. and the total energy $E_0^{\text{EHD}} = -1$ a.u..
This value corresponds to the energy of two separate hydrogen atoms.
We report the exact spectral function in Fig. \ref{fig:extHubbard_G1G3_2B_Exqp_Exact_H2} for various values of the intersite distance $R$.
These $R$ values represent different parts of the exact PES in Fig.~\ref{fig:extHubbard_gal_migdal_multiplot_3}, i.e.,  a value at which the total energy is 1) positive ($R=0.1$ a.u.); 2) negative with a negative slope ($R=1$ a.u.); 3) minimal ($R=1.44$ a.u.); 4) negative with a positive slope ($R=2.7$ a.u.); 5) close to the dissociation limit ($R=7$ a.u.). 
The chemical potential is set at $\mu=-(\epsilon^{\text{QP}}_{b}+\epsilon^{\text{QP}}_{a})/2$, with $\epsilon^{\text{QP}}_{b}$ and $\epsilon^{\text{QP}}_{a}$ the exact removal and addition QP energies, respectively.
It is also interesting to investigate the HOMO-LUMO gap of the EHD as a function of $R$ 
since this quantity is often difficult to reproduce by approximate self-energies especially in the the dissociation limit.
The exact HOMO-LUMO gap of the EHD, $\Delta^{\text{EHD}}_g$, is derived in App.~\ref{App:exact}) and it is given by
\begin{equation}   
\Delta^{\text{EHD}}_g = -2t+w+\sqrt{16t^2+(U-w)^2}.
\end{equation}
Its evolution as a function of $R$ is reported in Fig.~\ref{Fig:bandgaps}.
We see that the gap decreases by stretching the dimer and in the dissociation limit becomes equal to the onsite interaction parameter $U=5/8$ a.u. since both $t$ and $w$ vanish in the dissociation limit (see Fig.~\ref{fig:bondlengthRparaDep}).
\subsection{$(3,1)$-MCDE}
Solving the $(3,1)$-MCDE for the $2h1e$ and $2e1h$ channels of the 3-GF requires the corresponding 3-GF in the HF approximation.
The one-particle Hartree-Fock self-energy of the EHD is diagonal in the bonding/antibonding basis.
The diagonal elements are given by
\begin{equation}
	\Sigma^{HF}_{b\sigma;b\sigma}=\frac{U+w}{2};\;\;\;\;\;\; \Sigma^{HF}_{a\sigma;a\sigma}=\frac{U+3w}{2}.
 \label{Eqn:SHF-GHF}
\end{equation}
We can then obtain the HF 3-GF from equations (\ref{eq:g01im})-(\ref{eq:g03im}).
Its non-zero elements are given by
\begin{align}
\label{eq:ghf_ext_hubbard_888_2}
	G^{\text{HF}}_{b\sigma;b\sigma}&=\frac{1}{\omega-(\epsilon_0-t+\frac{w+U}{2})-\ii \eta},
	\\
	\label{eq:ghf_ext_hubbard_888}
	G^{\text{HF}}_{a\sigma;a\sigma}&=\frac{1}{\omega-(\epsilon_0+t+\frac{3w+U}{2})+\ii \eta},\\	
	G^{\text{HF}}_{b\downarrow b\uparrow a\sigma;b\downarrow b\uparrow a\sigma}&=
	\frac{1}{\omega-(\epsilon_0-3t+\frac{U-w}{2})-\ii \eta},\\
 \label{eq:ghf_ext_hubbard_888_end}
	G^{\text{HF}}_{a\downarrow a\uparrow b\sigma;a\downarrow a\uparrow b\sigma}&=\frac{1}{\omega-(\epsilon_0+3t+\frac{5w+U}{2})+\ii \eta}.
\end{align}
The non-zero elements of the 3-SE of the $(3,1)$-MCDE for the EHD are given by
\begin{align}
\Sigma_{3,a\downarrow a\uparrow b\sigma;a\downarrow a\uparrow b\sigma} & =-w
\\
\Sigma_{3,b\downarrow b\uparrow a\sigma;b\downarrow b\uparrow a\sigma} &= w
\\
\Sigma_{3,a\downarrow;b\downarrow b\uparrow a\uparrow} & = \Sigma_{3,b\downarrow;a\downarrow a\uparrow b\uparrow} 
= \frac{U-w}{2}
\\
\Sigma_{3,b\downarrow b\uparrow a\uparrow;a\downarrow} & = \Sigma_{3,a\downarrow a\uparrow b\uparrow;b\downarrow} 
= \frac{U-w}{2}
\\
\Sigma_{3,a\uparrow;b\downarrow b\uparrow a\downarrow} & = 
\Sigma_{3,b\uparrow;a\downarrow a\uparrow b\downarrow} = \frac{w-U}{2}
\\
\Sigma_{3,b\downarrow b\uparrow a\downarrow;a\uparrow} & = 
\Sigma_{3,a\downarrow a\uparrow b\downarrow;b\uparrow} = \frac{w-U}{2}
\end{align}
which, by construction, are all static.
We can now solve the $(3,1)$-MCDE for the $2h1e$ and $2e1h$ channels of the 3-GF.
The result is reported in Appendix \ref{App:exact}.
Here we just report the head of this 3-GF since it corresponds to the 1-GF which is the quantity we need to calculate the ground-state energy and the spectral function.
In the bonding/antibonding basis it is given by

\begin{align}
	\label{eq:gMCDE_ext_hubbard_888_2}
	G^{\text{MCDE}}_{b\sigma;b\sigma}&=\frac{\frac{1}{2}\left[1-\frac{4t+w}{g}\right]}{\omega-(\epsilon_0+t+\frac{2w+U+g}{2})+\ii\eta}\nonumber\\
 &+\frac{\frac{1}{2}\left[1+\frac{4t+w}{g}\right]}{\omega-(\epsilon_0+t+\frac{2w+U-g}{2})-\ii\eta},\\
 \label{eq:gMCDE_ext_hubbard_888}
	G^{\text{MCDE}}_{a\sigma;a\sigma}&=\frac{\frac{1}{2}\left[1+\frac{4t+w}{g}\right]}{\omega-(\epsilon_0-t+\frac{2w+U+g}{2})+\ii\eta} \nonumber\\
 &+\frac{\frac{1}{2}\left[1-\frac{4t+w}{g}\right]}{\omega-(\epsilon_0-t+\frac{2w+U-g}{2})-\ii\eta}.
\end{align}
with $g=\sqrt{16t^2+8tw+2w^2-2wU+U^2}$. 

%
\subsection{Second Born and $GW$}
We will compare the MCDE results to those obtained within the standard 2B and $GW$ approaches.
Therefore, we report here the 1-GFs for the EHD that correspond to these two approximations.
As mentioned before, for the case of 2B, it suffices to set $\Sigma^{c,\text{3p}}=0$ and solve the $(3,1)$-MCDE.
This will yield a 3-GF of which the head corresponds to the 1-GF within the second Born approximation.
To be precise it corresponds to the solution of Eq.~\eqref{Eq:oneshot2B}.
We get the following bonding/antibonding components of the 1-GF,
\begin{align}
G^{\text{2B}}_{b\sigma;b\sigma}&=\frac{\frac{1}{2}\left[1-\frac{4t+2w}{k}\right]}{\omega-(\epsilon_0+t+\frac{3w+U+k}{2})+\ii \eta}\\
&+\frac{\frac{1}{2}\left[1+\frac{4t+2w}{k}\right]}{\omega-(\epsilon_0+t+\frac{3w+U-k}{2})-\ii \eta}\nonumber\\
G^{\text{2B}}_{a\sigma;a\sigma}&=\frac{\frac{1}{2}\left[1+\frac{4t+2w}{k}\right]}{\omega-(\epsilon_0-t+\frac{w+U+k}{2})+\ii\eta}\\
&+\frac{\frac{1}{2}\left[1-\frac{4t+2w}{k}\right]}{\omega-(\epsilon_0-t+\frac{w+U-k}{2})-\ii \eta}\nonumber
\end{align}
with $k=\sqrt{16t^2+16tw+5w^2-2wU+U^2}$.

In the case of $GW$ we obtain.
\begin{align}
\label{Eqn:GbbGW}
G^{\textit{GW}}_{b\sigma;b\sigma}
&=\frac{\frac{1}{2}\left[1-\frac{h +2t +w }{d}\right]}{\omega-(\epsilon_0 + \frac{h}{2} + \frac{2w+U+d}{2})+\ii\eta}\nonumber\\
&+\frac{\frac{1}{2}\left[1+\frac{h +2t +w }{d}\right]}{\omega- (\epsilon_0 + \frac{h}{2} + \frac{2w+U-d}{2})-\ii\eta},\\
\label{Eqn:GaaGW}
G^{\textit{GW}}_{a\sigma;a\sigma}
&=\frac{\frac{1}{2}\left[1+\frac{h +2t +w }{d}\right]}{\omega-(\epsilon_0 - \frac{h}{2} + \frac{2w+U+d}{2} )+\ii\eta}\nonumber\\
&+\frac{\frac{1}{2}\left[1-\frac{h +2t +w }{d}\right]}{\omega- (\epsilon_0 - \frac{h}{2} + \frac{2w+U-d}{2})-\ii\eta}.
\end{align}
with $h=\sqrt{(2t+w)^2+2(U-w)(2t+w)}$ and $d=\sqrt{h^2+2h(2t+w)+(2t+w)^2+2(2t+w)(w-U)^2/h}$.
The details of the calculation are given in Appendix \ref{app:gw}.
\subsection{Comparison}
\subsubsection{Potential energy surface}
In Fig.~\ref{fig:extHubbard_gal_migdal_multiplot_3} we report a comparison of the potential energy surface of the EHD obtained from the various approximations to the 1-GF.
We observe that the $(3,1)$-MCDE yields a PES that is in overall good agreement with the exact one, especially considering that the PES  of the H$_2$ molecule is a challenging test case for many-body approximations.~\cite{Caruso_2013,Olsen_2014}
In particular, the $(3,1)$-MCDE gives a good description of the energy around the equilibrium distance (R = 1.44 a.u.) and it correctly yields the exact limit at $R\rightarrow \infty$.
Thanks to the analyticity of the EHD we can indeed confirm that the PES corresponding 
to the exact solution of the EHD as well as those obtained within the $(3,1)$-MCDE and 2B all yield the value $-1$ a.u. in the dissociation limit. This value corresponds to the energy of two separate sites with one electron each.
However, the $(3,1)$-MCDE overestimates the energy in the region between the equilibrium distance and the dissociation limit.
The PES obtained within the 2B approximation is very similar to the one obtained within the $(3,1)$-MCDE, although slightly worse. Instead, although $GW$ also accurately reproduces the energies around the equilibrium distance, it fails to correctly describe the energy in the dissociation limit. In this limit the $GW$ energy is equal to $-0.69$ a.u. compared to $-1$ a.u. for the exact solution.

It is worth noting that the PES of the EHD can also be obtained in a different manner.
Since, according to Eq.~\eqref{Eq:linkEHD-HD}, the energies of the standard and extended Hubbard dimers only differ by the intersite interaction $w$, one can use the GM formula in Eq.~\eqref{Eq:GM} with the 1-GF of the standard Hubbard dimer, and then obtain the energy for the EHD from Eq.~\eqref{Eq:linkEHD-HD}.
This is an interesting alternative because the $(3,1)$-MCDE yields the exact 1-GF for the Hubbard dimer at 1/2 filling~\cite{riva_prl}.
Therefore, we obtain the exact PES of the EHD with this approach.
Finally, we note that one could even obtain the exact PES of the EHD within the 2B approximation since Eq.~\eqref{Eq:oneshot2B}  yields the exact 1-GF for the standard Hubbard dimer.~\cite{riva_prb}
The exact PES for the EHD can then again be obtained from the GM formula and Eq.~\eqref{Eq:linkEHD-HD}.
%
\subsubsection{Spectral function}
In Fig.~\ref{fig:extHubbard_G1G3_2B_Exqp_Exact_H2} we report a comparison of the spectral functions obtained within the various approximations for several values of $R$.
We see that the $(3,1)$-MCDE spectra are in very good agreement with the exact spectral functions over the whole range of $R$ values.
In particular, the satellite positions are in excellent agreement with the exact results even for large values of $R$ where electron correlation is important.
Although the PES obtained within the 2B and $(3,1)$-MCDE approximations are very similar, the 2B spectral functions are clearly worse than those obtained with the $(3,1)$-MCDE. Again, the worst results are obtained with the $GW$ approximation.
Both 2B and $GW$ are unable to correctly describe the satellite peaks in the spectra and even the QP peaks are not well described for large values of $R$.
\subsubsection{HOMO-LUMO gaps}
Finally, we report the HOMO-LUMO gap obtained with the various approximations as a function of the bond length $R$ in Fig.~\ref{Fig:bandgaps}.
The curves show a very similar trend to those obtained for the PES.
The best results are again obtained with the $(3,1)$-MCDE with the 2B gaps being slightly worse over the whole $R$ range.
Both $(3,1)$-MCDE and 2B yield the exact gap ($\Delta_g=5/8$ a.u.) in the limit $R\rightarrow\infty$.
Finally, $GW$ is unable to yield accurate gaps for large $R$.
Moreover, in the dissociation limit the gap obtained within $GW$ vanishes instead of remaining finite as it should. \cite{Rom09-1,Rom12,Dis15}

%
%
\subsection{Analysis}
It is instructive to inspect the differences between the approximate 3-SE of the $(3,1)$-MCDE and the exact 3-SE.
Thanks to the analyticity of the EHD, the exact 3-SE can be obtained from the exact 3-GF and the HF 3-GF by solving Eq.~\eqref{eq:gDyson}.
Its non-zero elements in the bonding/antibonding basis are reported in Table \ref{tab:exactMCDEcomparison}
and compared to those of the $(3,1)$-MCDE.
As expected, the two self-energies become equal in the limit $w\rightarrow 0$ since this limit corresponds to the standard Hubbard dimer for which the $(3,1)$-MCDE is exact.~\cite{riva_prl}

The most striking difference between the two 3-SE is in their one-particle channels:  
it is non-zero for the exact 3-SE but vanishes in the $(3,1)$-MCDE.
Since the $(3,1)$-MCDE is exact up to second order in the interaction~\cite{riva_prl,riva_prb} it means that these elements should be at least of third order in the interaction.
Indeed, using the expansion $\frac{1}{c}\approx \frac{1}{4t}\left(1-\frac{(w-U)^2}{32t^2}\right)$, it can be verified that the leading order of these elements is of third order in the interaction.
The difference between the two 3-SE can be traced back to the calculation of the $G_3^{\text{HF}}$.
In the $(3,1)$-MCDE the HF 3-GF is constructed from the HF 1-GF obtained from the HF self-energy calculated from the HF 1-GF, 
i.e., as $\Sigma_1^{\text{HF}}[G_1^{\text{HF}}]$.
Instead, in Eq.\eqref{Eq:Sigma1decomp} the static part of the self-energy is $\Sigma_1^{\text{HF}}[G_1]$, i.e., 
the HF self-energy is calculated from the exact 1-GF.
For the EHD the HF self-energy as a functional of the exact 1-GF reads 
\begin{eqnarray}\label{Eqn:exactHFa}
	\Sigma^{HF}_{a\sigma;a\sigma}[G^\text{EHD}_1]&=&\frac{3w+U}{2}+w\left(\frac{2t}{c}-\frac12\right),\\
	\label{Eqn:exactHFb}\Sigma^{HF}_{b\sigma;b\sigma}[G^\text{EHD}_1]&=&\frac{w+U}{2}-w\left(\frac{2t}{c}-\frac12\right),
\end{eqnarray}
which should be compared to Eq.~\eqref{Eqn:SHF-GHF} for the HF self-energy as a functional of the HF 1-GF.
The difference between the two self-energies becomes increasingly negligible in the dissociation limit, since in this limit $w\rightarrow 0$.
%
\begin{figure*}[t]
	\centering
	\includegraphics[scale=\scale]{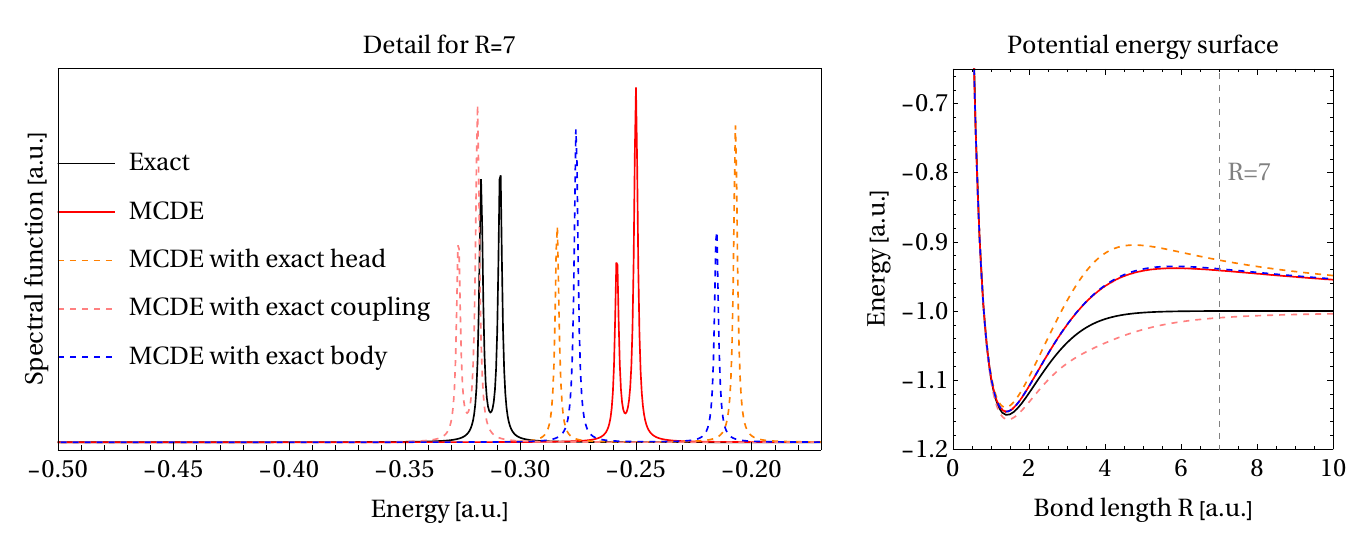}
	\caption{Left panel: the removal part of the spectral function $A(\omega)$ of the extended Hubbard dimer at bond length $R=7$ a.u. obtained with various hybrid self-energies in which part of the $(3,1)$-MCDE self-energy is replaced by the corresponding part of the exact self-energy. The results are compared to the exact spectral function.  Right panel: The ground-state energy as a function of the bond length $R$ obtained within the same approximations and compared to the exact potential-energy surface.}
	\label{fig:multiplotHFEX}
\end{figure*}

We have analyzed the impact on the results of substituting the one-particle channel of the 3-SE within the $(3,1)$-MCDE approximation with the one-particle channel of the exact 3-SE. 
We note that this is equivalent to using Eqs.~\eqref{Eqn:exactHFa} and ~\eqref{Eqn:exactHFb} instead of Eq.~\eqref{Eqn:SHF-GHF} to obtain the HF QP energies that enter the 1-particle channel of the effective Hamiltonian.
We report the PES and spectral functions thus obtained in Fig.~\ref{fig:multiplotHFEX}.
We clearly see that both the PES and spectral functions deteriorate, which indicates that improving only the static part of the 1-SE does not necessarily yield improved results.
We note that both spectral functions fulfill the sum rule~\cite{vonBarth_PRB96}
%
\begin{equation}
\int^{\infty}_{-\infty} d\omega \omega A_{ii}(\omega)=\epsilon_i^{\text{HF}}
\end{equation}
where $i=b,a$ and $\epsilon_i^{\text{HF}}$ are the HF quasiparticle energies obtained from the exact 1-GF in the case of the MCDE spectral function obtained from $G^{\text{HF}}_1[G_1]$, and the HF quasiparticle energies obtained from the HF 1-GF in the case of the MCDE spectral function obtained from $G^{HF}_1[G^{\text{HF}}_1]$.

This analysis raises the question if, instead, the results could be improved by replacing the three-particle channel of the $(3,1)$-MCDE by the three-particle channel of the exact 3-SE, or by a similar substitution for the coupling of the one- and three-particle channels.
In Fig. \ref{fig:multiplotHFEX} we report the PES and spectral functions obtained with these hybrid self-energies.
We observe that substituting the couplings of the 3-SE yields an improved PES as well as improved spectral functions, while the substitution of the three-particle channel worsens the results.
The leading order of the difference between the couplings is of second order in the interaction, while the leading order of the difference between the exact and approximate 1-particle and 3-particle channels are both of third order in the interaction.
Moreover, at least for the EHD, the difference between the exact and approximate 1-particle and 3-particle channels are very similar.
This indicates that there is an error cancellation between the 1-particle and 3-particle channels of 3-SE of the $(3,1)$-MCDE.
In conclusion, the above analysis shows that the most efficient way to improve the self-energy of the $(3,1)$-MCDE is by improving the coupling terms, e.g., by including the second-order contributions in the interaction.\cite{Sch78,Sch83,Nie84,Deleuze_96}. Moreover, the approximations to the 1-particle channel (head) and the 3-particle channel (body) of the $(3,1)$-MCDE self-energy should be improved in a consistent manner since for both channels the missing correlation has the same leading order in the interaction.

%
\begin{table}[t]	
\caption{
	Analytic expressions for the exact and $(3,1)$-MCDE self-energies in the bonding/antibonding basis. 
}
\label{tab:exactMCDEcomparison}
	\begin{center}
		\begin{tabular}[t]{ccc}
			$\Sigma_3$ & Exact & $(3,1)$-MCDE \\
			\hline\hline
			$\Sigma_{3,a\sigma;a\sigma}$&
			$-w\left(\frac{1}{2} - \frac{2t}{c}\right)$
			& 0
			\\
			$\Sigma_{3,b\sigma;b\sigma}$&
			$w\left(\frac{1}{2} - \frac{2t}{c}\right)$
			& 0
			\\
			\hline
			$\Sigma_{3,a\downarrow a\uparrow b\sigma;a\downarrow a\uparrow b\sigma}$&
			$-w \left(\frac{3}{2}-\frac{2 t}{c}\right)$
			& $-w$ 
			\\
			$\Sigma_{3,b\downarrow b\uparrow a\sigma;b\downarrow b\uparrow a\sigma}$&
			$w \left(\frac{3}{2} - \frac{2 t}{c}\right)$
			& $w$ 
			\\
			\hline
			$\Sigma_{3,a\downarrow;b\downarrow b\uparrow a\uparrow}$&\multirow{4}{*}{$\frac{(U-w)}{2} \left(1+\frac{w}{c}\right)$}&\multirow{4}{*}{$\frac{U-w}{2}$}\\
			$\Sigma_{3,b\downarrow;a\downarrow a\uparrow b\uparrow}$& & \\
			$\Sigma_{3,b\downarrow b\uparrow a\uparrow;a\downarrow}$& &  \\
			$\Sigma_{3,a\downarrow a\uparrow b\uparrow;b\downarrow}$ & &
			 \\
			\hline
			$\Sigma_{3,a\uparrow;b\downarrow b\uparrow a\downarrow}$&\multirow{4}{*}{$\frac{(w-U)}{2} \left(1+\frac{w}{c}\right)$}&\multirow{4}{*}{$\frac{w-U}{2}$}\\
			$\Sigma_{3,b\uparrow;a\downarrow a\uparrow b\downarrow}$& &  \\
			$\Sigma_{3,b\downarrow b\uparrow a\downarrow;a\uparrow}$& &  \\
			$\Sigma_{3,a\downarrow a\uparrow b\downarrow;b\uparrow}$ & &
			\\
			\hline\hline
		\end{tabular}
	\end{center}	
\end{table}
\section{Conclusions\label{sec:conclusions}}
In this work we scrutinized the quality of the multichannel Dyson equation (MCDE), recently introduced by some of us, on the extended Hubbard dimer. 
This model is closely related to the H$_2$ molecule and it allows us to study the various correlation regimes along the H$_2$ dissociation curve. 
Since the model is analytically solvable, it represents an important benchmark for our new method. 
We showed that the $(3,1)$-MCDE outperforms both the second Born and $GW$ approximations, two standard many-body approaches, for all properties studied in this work i.e., the PES, spectral functions and HOMO-LUMO gaps.
The spectral functions obtained within the $(3,1)$-MCDE also compare very well to the exact results.
The $(3,1)$-MCDE also correctly describes the PES around the equilibrium distance as well as in the dissociation limit, where the $(3,1)$-MCDE is exact.
However, in the region before full dissociation the $(3,1)$-MCDE energies are overestimated.
We obtain similar results for the HOMO-LUMO gap, i.e., around the equilibrium distance the $(3,1)$-MCDE correctly describes the gap and in the dissociation limit the $(3,1)$-MCDE yields the exact gap.
In the region before full dissociation the $(3,1)$-MCDE underestimates the gaps.
These results demonstrate that the $(3,1)$-MCDE is a promising approach for both weakly and strongly correlated systems but also that there is still room for improvement.
Finally, we showed that such improvements could be most easily obtained by improving the quality of the coupling in the three-body self-energy of the $(3,1)$-MCDE.

\textit{Acknowledgment:}
We thank the French “Agence Nationale de la Recherche (ANR)” for financial support (Grant Agreements No. ANR-19-CE30-0011 and No. ANR-22-CE30-0027).

\appendix
\section{Exact results\label{App:exact}}	
To obtain the exact $G_3$ of the extended Hubbard dimer at half-filling (2 electrons), we have to evaluate the following components of $G_3$,
{\footnotesize
\begin{align}
	\label{eq:g2spectrumformula}
	G^{EHD}_{i;j}(\omega)&=\sum_{n}\frac{\bra{\Psi_0^{N=2}}c_{i}\ket{\Psi^{N=3}_n}\bra{\Psi^{N=3}_n}c_{j}^\dagger\ket{\Psi_0^{N=2}}}{\omega-(E^{N=3}_n-E_{0}^{N=2})+\ii \eta}\nonumber\\
	&+
	\frac{\bra{\Psi_0^{N=2}}c_{i}^\dagger\ket{\Psi^{N=1}_n}\bra{\Psi^{N=1}_n}c_{j}\ket{\Psi_0^{N=2}}}{\omega-(E_{0}^{N=2}-E^{N=1}_n)-\ii \eta},
	\\
	\label{eq:g2spectrumformula_3part}
	G^{EHD}_{i jl;m o k}(\omega)
	&=
	\sum_{n}\frac{\bra{\Psi_0^{N=2}}c_{l}^\dagger c_{j}c_{i}\ket{\Psi^{N=3}_n}\bra{\Psi^{N=3}_n}c_{m}^\dagger c_{o}^\dagger c_{k}\ket{\Psi_0^{N=2}}}{\omega-(E^{N=3}_n-E_{0}^{N=2})+\ii \eta}\nonumber\\
	&+
	\frac{\bra{\Psi_0^{N=2}}c_{i}^\dagger c_{j}^\dagger c_{l}\ket{\Psi^{N=1}_n}\bra{\Psi^{N=1}_n}c_{k}^\dagger c_{o}c_{m}\ket{\Psi_0^{N=2}}}{\omega-(E_{0}^{N=2}-E^{N=1}_n)-\ii \eta},
	\\
	\label{eq:g2spectrumformula_3coupling1}
	G^{EHD}_{i;m o k}(\omega)
	&=
	\sum_{n}\frac{\bra{\Psi_0^{N=2}}c_{i}\ket{\Psi^{N=3}_n}\bra{\Psi^{N=3}_n}c_{m}^\dagger c_{o}^\dagger c_{k}\ket{\Psi_0^{N=2}}}{\omega-(E^{N=3}_n-E_{0}^{N=2})+\ii \eta}\nonumber\\
	&+
	\frac{\bra{\Psi_0^{N=2}}c_{i}^\dagger\ket{\Psi^{N=1}_n}\bra{\Psi^{N=1}_n}c_{k}^\dagger c_{o}c_{m}\ket{\Psi_0^{N=2}}}{\omega-(E_{0}^{N=2}-E^{N=1}_n)-\ii \eta},
	\\
	\label{eq:g2spectrumformula_3coupling2}
	G^{EHD}_{i jl;m}(\omega)
	&=
	\sum_{n}\frac{\bra{\Psi_0^{N=2}}c_{l}^\dagger c_{j}c_{i}\ket{\Psi^{N=3}_n}\bra{\Psi^{N=3}_n}c_{m}^\dagger\ket{\Psi_0^{N=2}}}{\omega-(E^{N=3}_n-E_{0}^{N=2})+\ii \eta}\nonumber\\
	&+
	\frac{\bra{\Psi_0^{N=2}}c_{i}^\dagger c_{j}^\dagger c_{l}\ket{\Psi^{N=1}_n}\bra{\Psi^{N=1}_n}c_{m}\ket{\Psi_0^{N=2}}}{\omega-(E_{0}^{N=2}-E^{N=1}_n)-\ii \eta},
\end{align}
}
where $\ket{\Psi^{N=k}_{n}}$ and $E_{n}^{N=k}$ refer to the $n$th eigenvector and corresponding eigenvalue, respectively, of the $k$-electron system. 
 
At half-filling the ground-state wavefunction in the bonding/antibonding (b/a) basis is given by 
\begin{equation}
\Psi^{N=2}_0= A|b\uparrow b\downarrow\rangle+ B|a\uparrow a\downarrow\rangle
\end{equation}
where $|R\sigma\,S\sigma'\,...\rangle$ is a Slater determinant with on its diagonal $R\sigma$, $S\sigma'$, and so on. For a one-electron system $|R\sigma\rangle$ thus indicates a one-electron spinorbital.
\begin{align}
A&=\frac{1+\frac{4t}{c+(w-U)}}{a}
\\
B&=\frac{1-\frac{4t}{c+(w-U)}}{a}
\\
a&=\sqrt{2\left(\frac{16t^2}{(c+(w-U))^2}+1\right)}
\\
c&=\sqrt{16t^2+(w-U)^2}.
\end{align}
The corresponding ground-state energy is 
\begin{equation}
E^{N=2}_{0}=2\epsilon_0+\frac{w+U-c}{2}
\end{equation}

The $3$-electron system has the following eigenvectors and eigenvalues: 
\begin{align}
\Psi^{N=3}_0&=|	b\uparrow b\downarrow a\uparrow\rangle,\, E^{N=3}_{0}=3\epsilon_0+2w+U-t,\\
\Psi^{N=3}_1&=|	b\uparrow b\downarrow a\downarrow\rangle,\, E^{N=3}_{1}=3\epsilon_0+2w+U-t,\\
	\Psi^{N=3}_2&=|	b\uparrow a\uparrow a\downarrow \rangle, E^{N=3}_2=3\epsilon_0+2w+U+t\\
 \Psi^{N=3}_3&=|	b\downarrow a\uparrow a\downarrow \rangle, E^{N=3}_3=3\epsilon_0+2w+U+t.
\end{align}

Finally, the one-particle eigenvectors and eigenenergies are 
\begin{align}
\Psi^{N=1}_0&=|	b\uparrow \rangle,\, E^{N=1}_{0}=\epsilon_0-t,\\
\Psi^{N=1}_1&=|	 b\downarrow \rangle,\, E^{N=1}_{1}=\epsilon_0-t,\\
	\Psi^{N=1}_2&=|	a\uparrow  \rangle, E^{N=1}_2=\epsilon_0+t\\
 \Psi^{N=1}_3&=|	a\downarrow \rangle, E^{N=1}_3=\epsilon_0+t.
\end{align}
Note that for the 1- and 3-electron case we chose as ground-state the spin-up configuration for the unpaired electron, but of course we could have equally chosen the spin-down configuration.

Using  these ingredients in Eq.~ (\ref{eq:g2spectrumformula}) we can compute the exact $G_3$, which reads
\begin{align}
\label{eq:g_ext_hubbard_888_2}
	G^{\text{EHD}}_{b\sigma;b\sigma}&=\frac{B^2}{\omega-(\epsilon_0+t+\frac{3w+U+c}{2})+\ii \eta}\nonumber\\
 &+\frac{A^2}{\omega-(\epsilon_0+t+\frac{w+U-c}{2})-\ii \eta},
	\\
	\label{eq:g_ext_hubbard_888}
	G^{\text{EHD}}_{a\sigma;a\sigma}&=\frac{A^2}{\omega-(\epsilon_0-t+\frac{3w+U+c}{2})+\ii \eta}\nonumber\\
 &+\frac{B^2}{\omega-(\epsilon_0-t+\frac{w+U-c}{2})-\ii \eta},\\
 \label{eq:g_ext_hubbard_888_end}
	G^{\text{EHD}}_{b\downarrow b\uparrow a\sigma;b\downarrow b\uparrow a\sigma}&=
	\frac{B^2}{\omega-(\epsilon_0-t+\frac{3w+U+c}{2})+\ii \eta}\nonumber\\
 &+\frac{A^2}{\omega-(\epsilon_0-t+\frac{w+U-c}{2})-\ii \eta},\\
	G^{\text{EHD}}_{a\downarrow a\uparrow b\sigma;a\downarrow a\uparrow b\sigma}&=\frac{A^2}{\omega-(\epsilon_0+t+\frac{3w+U+c}{2})+\ii \eta}\nonumber\\
 &+\frac{B^2}{\omega-(\epsilon_0+t+\frac{w+U-c}{2})-\ii \eta},\\
	G^{\text{EHD}}_{b\downarrow b\uparrow a\uparrow;a\downarrow }&=G^{\text{EHD}}_{a\downarrow;b\downarrow b\uparrow a\uparrow}=\frac{-AB}{\omega-(\epsilon_0-t+\frac{3w+U+c}{2})+\ii \eta}\nonumber\\
 &+\frac{AB}{\omega-(\epsilon_0-t+\frac{w+U-c}{2})-\ii \eta},\\
	G^{\text{EHD}}_{b\downarrow b\uparrow a\downarrow;a\uparrow }&=G^{\text{EHD}}_{a\uparrow;b\downarrow b\uparrow a\downarrow}=\frac{AB}{\omega-(\epsilon_0-t+\frac{3w+U+c}{2})+\ii \eta}\nonumber\\
 &+\frac{-AB}{\omega-(\epsilon_0-t+\frac{w+U-c}{2})-\ii \eta},\\
	G^{\text{EHD}}_{a\downarrow a\uparrow b\uparrow;b\downarrow }&=G^{\text{EHD}}_{b\downarrow;a\downarrow a\uparrow b\uparrow}=\frac{-AB}{\omega-(\epsilon_0+t+\frac{3w+U+c}{2})+\ii \eta}\nonumber\\
 &+\frac{AB}{\omega-(\epsilon_0+t+\frac{w+U-c}{2})-\ii \eta},\\
	G^{\text{EHD}}_{a\downarrow a\uparrow b\downarrow;b\uparrow }&=G^{\text{EHD}}_{b\uparrow;a\downarrow a\uparrow b\downarrow}=\frac{AB}{\omega-(\epsilon_0+t+\frac{3w+U+c}{2})+\ii \eta}\nonumber\\
 &+\frac{-AB}{\omega-(\epsilon_0+t+\frac{w+U-c}{2})-\ii \eta}.
\end{align}

We notice that the poles $\epsilon_a^{\text{QO}}=\epsilon_0-t+\frac{3w+U+c}{2}$ and $\epsilon_b^{\text{QP}}=\epsilon_0+t+\frac{w+U-c}{2}$ are the electron affinity (EA) and the ionization potential (IP) energies, respectively. The exact HOMO-LUMO gap is hence  $\Delta^{\text{EHD}}_g=\epsilon_a^{\text{QP}}-\epsilon_b^{\text{QP}}=-2t+w+c$.

The exact three-body self-energy (at 1/2 filling) is obtained as $\Sigma^{\text{EHD}}_3=[G^{\text{HF}}_{3}]^{-1}-[G^{\text{EHD}}_{3}]^{-1}$. 
The non-zero elements of the exact self-energy are displayed in Table \ref{tab:exactMCDEcomparison} and compared with the MCDE $\Sigma_3$. The $[G^{\text{HF}}_{3}]^{-1}$ is reported in Equations (\ref{eq:ghf_ext_hubbard_888_2})-(\ref{eq:ghf_ext_hubbard_888_end}).
It is worth noting that $\Sigma^{\text{EHD}}_{3}$ is $\omega$-independent.
\section{MCDE 3-body Green's function}
Besides the elements given in (Eqs \ref{eq:gMCDE_ext_hubbard_888_2})-(\ref{eq:gMCDE_ext_hubbard_888}), the remaining elements of the 3-GF within the $(3,1)$-MCDE are as follows

\begin{align}
G^{\text{MCDE}}_{b\downarrow b\uparrow a\sigma;b\downarrow b\uparrow a\sigma}&=
	\frac{\frac{1}{2}\left[1-\frac{4t+w}{g}\right]}{\omega-(\epsilon_0-t+\frac{2w+U+g}{2})+\ii\eta}\nonumber\\
 &+\frac{\frac{1}{2}\left[1+\frac{4t+w}{g}\right]}{\omega-(\epsilon_0-t+\frac{2w+U-g}{2})-\ii\eta},\\
	G^{\text{MCDE}}_{a\downarrow a\uparrow b\sigma;a\downarrow a\uparrow b\sigma}&=\frac{\frac{1}{2}\left[1+\frac{4t+w}{g}\right]}{\omega-(\epsilon_0+t+\frac{2w+U+g}{2})+\ii\eta}\nonumber\\
 &+\frac{\frac{1}{2}\left[1-\frac{4t+w}{g}\right]}{\omega-(\epsilon_0+t+\frac{2w+U-g}{2})-\ii\eta},\\
	G^{\text{MCDE}}_{b\downarrow b\uparrow a\uparrow;a\downarrow }&=G^{\text{MCDE}}_{a\downarrow;b\downarrow b\uparrow a\uparrow}=\frac{\frac{U-w}{2}}{\omega-(\epsilon_0-t+\frac{2w+U+g}{2})+\ii \eta}\nonumber\\
 &+\frac{\frac{w-U}{2}}{\omega-(\epsilon_0-t+\frac{2w+U-g}{2})-\ii \eta},\\
	G^{\text{MCDE}}_{b\downarrow b\uparrow a\downarrow;a\uparrow }&=G^{\text{MCDE}}_{a\uparrow;b\downarrow b\uparrow a\downarrow}=\frac{\frac{w-U}{2}}{\omega-(\epsilon_0-t+\frac{2w+U+g}{2})+\ii \eta}\nonumber\\
 &+\frac{\frac{U-w}{2}}{\omega-(\epsilon_0-t+\frac{2w+U-g}{2})-\ii \eta},\\
	G^{\text{MCDE}}_{a\downarrow a\uparrow b\uparrow;b\downarrow }&=G^{\text{MCDE}}_{b\downarrow;a\downarrow a\uparrow b\uparrow}=\frac{\frac{U-w}{2}}{\omega-(\epsilon_0+t+\frac{2w+U+g}{2})+\ii \eta}\nonumber\\
 &+\frac{\frac{w-U}{2}}{\omega-(\epsilon_0+t+\frac{2w+U-g}{2})-\ii \eta},\\
	G^{\text{MCDE}}_{a\downarrow a\uparrow b\downarrow;b\uparrow }&=G^{\text{MCDE}}_{b\uparrow;a\downarrow a\uparrow b\downarrow}=\frac{\frac{w-U}{2}}{\omega-(\epsilon_0+t+\frac{2w+U+g}{2})+\ii \eta}\nonumber\\
 &+\frac{\frac{U-w}{2}}{\omega-(\epsilon_0+t+\frac{2w+U-g}{2})-\ii \eta}.
\end{align}
 
\section{One-shot $GW$}
	\label{app:gw}
The screened Coulomb potential $W$ is calculated as $W=[1-v_cP]^{-1}v_c$, where $v_c$ is the bare Coulomb potential, and $P=\frac{1}{2\pi \ii}\int\text{d}\omega^{\prime} G_1(\omega+\omega^\prime)G_1(\omega^\prime)$  the polarizability.
In general $W$ is a four-point function when projected in a basis set. However in the extended Hubbard dimer it depends only on two indices as follows,
\begin{align}
		{W}_{ijji}&=\sum_l M^{-1}_{il}\left[U\delta_{lj}+w(1-\delta_{lj})\right]
\end{align}
with
\begin{align}
M_{il}= \left\{\delta_{il}-\sum_a\left(U\delta_{ia}+w(1-\delta_{ia})\right)P_{alla}\right\}
\end{align}
where $P_{alla}=G_{al}G_{la}$. We will work in the site basis since it is more convenient than the bonding/antibonding basis for the present derivation. We build $P$ from the HF $G_1$, which in the site basis, reads  
	\begin{align}
	G^{\text{HF}}_{ij}&=\frac12\left[\frac{(-1)^{(i-j)}}{\omega-(t+\frac{3w+U}{2}+\epsilon_0)+\ii\eta}\right]
	\nonumber\\
	&+\frac12\left[\frac{1}{\omega-(-t+\frac{w+U}{2}+\epsilon_0)-\ii\eta}\right],	
	\end{align}
 with $i,j$ running over the site 1 and the site 2. The polarizability hence reads
	\begin{equation}
		P_{ijji}(\omega)=\frac{1}{2}\left[\frac{(-1)^{(i-j)}(2t+w)}{\omega^2-(2t+w)^2}\right].
	\end{equation}
	We thus arrive at the following expressions for $W$

	\begin{align}
		W_{1111}=W_{11}&=U+\frac{(U-w)^2(2t+w)}{\omega^2-(2t+w)^2-2(U-w)(2t+w)},\\
		W_{1221}=W_{12}&=w-\frac{(U-w)^2(2t+w)}{\omega^2-(2t+w)^2-2(U-w)(2t+w)}.
	\end{align}
	The first (static) term on the right-hand side of the above equations gives rise to the exchange term, which is already included at the level of the $G_1^{HF}$. We will concentrate hence only on the correlation (frequency-dependent) part $W^c=W-v_c$. The (correlation contribution to the) GW self-energy $\Sigma_{ij}^{\text{c},GW}(\omega)=\frac{\ii}{2\pi}\int d\omega' G_{ij}^{\text{HF}}(\omega+\omega')W^c_{ij}(\omega')e^{i\omega'\eta}$ reads
 \begin{align}
 \Sigma^{\text{c},GW}_{ij}(\omega)&=-\frac{1}{2}U\delta_{ij}+\frac{(-1)^{(i-j)}w-w}{4}+\frac{(U-w)^2(2t+w)}{4h} \nonumber\\
 &\times \Big[\frac{(-1)^{(i-j)}}{\omega-(\epsilon_0-t+(w+U)/2)+h-\ii\eta}\nonumber\\
 &+\frac{1}{\omega-h-(\epsilon_0+t+(3w+U)/2)+\ii\eta}\Big]
     \end{align}

In the bonding/antibonding basis the $GW$ self-energy reads
\begin{align}
	\Sigma^{\text{c},GW}_{bb}(\omega)&=\Sigma_{11}(\omega)+\Sigma_{12}(\omega),\\
	\Sigma^{\text{c},GW}_{aa}(\omega)&=\Sigma_{11}(\omega)-\Sigma_{12}(\omega).
\end{align} 
from which one can calculate the one-body Green's function in Eqs \eqref{Eqn:GbbGW}-\eqref{Eqn:GaaGW}.

	\bibliographystyle{apsrev}

\end{document}